\documentclass[iop,apj]{emulateapj}

\pdfoutput=1

\usepackage{graphicx}
\usepackage{aas_macros}
\usepackage{amsmath}
\usepackage{amssymb}
\usepackage{latexsym}
\usepackage{color}
\usepackage{url}
\usepackage{hyperref}
\usepackage{ulem}

\begin{document}

\newcommand{\figscl}{0.3}
\newcommand{\ihmpc}{\,$h$\,Mpc$^{-1}$}
\newcommand{\hmpc}{\,$h^{-1}$\,Mpc}
\newcommand{\hgpc}{\,$h^{-1}$\,Gpc}
\newcommand{\hmsun}{\,$h^{-1}\,M_{\odot}$}
\newcommand{\kny}{\,$k_{\rm{Ny}}$}

\shorttitle{Corner Modes in Initial Conditions}
\shortauthors{Falck, et al.}

\title{The Effect of Corner Modes in the Initial Conditions of Cosmological Simulations}
\author{B. Falck\altaffilmark{1}, N. McCullagh\altaffilmark{2}, M. C. Neyrinck\altaffilmark{2,3,4}, J. Wang\altaffilmark{5}, A. S. Szalay\altaffilmark{3}}
\altaffiltext{1}{Institute of Theoretical Astrophysics, University of Oslo, PO Box 1029 Blindern, N-0315, Oslo, Norway}
\altaffiltext{2}{Institute for Computational Cosmology, Department of Physics, Durham University, Durham DH1 3LE, UK}
\altaffiltext{3}{Department of Physics and Astronomy, Johns Hopkins University, 3400 N Charles St, Baltimore, MD 21218, USA}
\altaffiltext{4}{Institut d'Astrophysique de Paris, 98 bis boulevard Arago, 75014, Paris, France}
\altaffiltext{5}{National Astronomy Observatories, Chinese Academy of Science, Datun Road, Beijing, PR China}

\begin{abstract}

In view of future high-precision large-scale structure surveys, it is important to quantify the percent and subpercent level effects in cosmological $N$-body simulations from which theoretical predictions are drawn. One such effect involves deciding whether to zero all modes above the one-dimensional Nyquist frequency, the so-called ``corner'' modes, in the initial conditions. We investigate this effect by comparing power spectra, density distribution functions, halo mass functions, and halo profiles in simulations with and without these modes. For a simulation with a mass resolution of $m_p \sim 10^{11}$\hmsun, we find that at $z>6$, the difference in the matter power spectrum is large at wavenumbers above $\sim 80$\% of \kny, reducing to below 2\% at all scales by $z\sim 3$. Including corner modes results in a better match between low- and high-resolution simulations at wavenumbers around the Nyquist frequency of the low-resolution simulation, but the effect of the corner modes is smaller than the effect of particle discreteness. The differences in mass functions are 3\% for the smallest halos at $z=6$ for the $m_p \sim 10^{11}$\hmsun\ simulation, but we find no significant difference in the stacked profiles of well-resolved halos at $z \leq 6$. Thus removing power at $|\mathbf{k}|>$\kny\ in the initial conditions of cosmological simulations has a small effect on small scales and high redshifts, typically below a few percent.

\end{abstract}


\section{Introduction}

The growth of structure in the Universe is a highly nonlinear process, making it difficult to calculate precise theoretical predictions for the late-time distribution of large-scale structure. Cosmological $N$-body simulations have revolutionized our understanding of large-scale structure in the Universe by numerically solving for the nonlinear gravitational collapse of matter, enabling us to both test and rule out cosmological theories and to calibrate analysis methods used in observations~\citep[for reviews, see][]{Efstathiou1985,Bertschinger1998}. As computer hardware and numerical algorithms have improved over the past few decades, cosmological simulations have been able to simulate larger volumes and higher resolutions and include non-standard cosmological models and complex baryonic physical processes, increasing the accuracy with which we can model the formation of structure in the Universe.

As future cosmological surveys push their measurements in order to achieve higher and higher precision (e.g. Euclid, WFIRST, DESI, LSST, and SKA), it is important that theoretical predictions taken from numerical simulations do not significantly contribute to the error in the derived cosmological parameters. This means both that the numerical codes must be pushed to achieve higher accuracy and that previously minor processes must be taken into account. Authors of numerical codes perform their own consistency checks on problems with known analytical solutions; nevertheless, small but measurable differences remain in full cosmological $N$-body simulations~\citep{Heitmann2008,Heitmann2010,Schneider2015}. Even if different numerical codes agreed, pushing simulations to achieve higher accuracy requires taking into account effects such as, for example, how the initial conditions are generated~\citep{Crocce2006,LHuillier2014}, the choice of starting redshift~\citep{Heitmann2010,Reed2013,McCullagh2016}, the force softening~\citep{Smith2014}, general relativistic effects~\citep{Christopherson2015,Thomas2015}, and baryonic effects~\citep{vanDaalen2011}.

Here we focus here on the generation of initial conditions. Many methods and publicly available codes exist~\citep[e.g.][]{Bertschinger1995,Pen1997,Bertschinger2001,Sirko2005,Crocce2006,Jenkins2010,Hahn2011,Jenkins2013}, some of which are designed to work with a specific $N$-body code. Generating initial conditions involves first calculating the spectrum of density perturbations at some high redshift, given a background cosmological model. If the primordial density fluctuations are Gaussian, which they are in the simplest models of inflation, then their statistics are fully specified by the power spectrum, $P(k)$. The cosmic variance from realization to realization consists of different sets of Fourier phases and amplitudes consistent with $P(k)$. The power spectrum can be written as the product of a power law spectrum (with power given by $n_s$, the spectral index after inflation) with random phases and the transfer function that represents the linear evolution of each mode. The transfer function can be calculated numerically (e.g. LINGER~\citep{Ma1995}, CMBFast~\citep{Seljak1996}, CAMB~\citep{Lewis2000}, or CLASS~\citep{Blas2011}), or calculated using approximate analytical expressions~\citep[e.g. ][]{Eisenstein1998}. 
This random phase realization of a given power spectrum must then be discretized to a grid of some resolution, depending on the requirements of the simulation. 
Dark matter particle positions and velocities at a (chosen) high redshift are obtained by perturbing the discretized initial power spectrum from either a uniform grid or a force-neutral ``glass'' configuration, according to first order~\citep[the Zel'dovich approximation, ][]{Zeldovich1970} or second order Lagrangian perturbation theory (2LPT). 

Each step in the generation of initial conditions involves choices that affect the numerical accuracy of the high redshift density field or power spectrum and which can propagate to lower redshifts. For example, the real space density field can either be calculated by taking the inverse Fourier transform of the Fourier space transfer function or by convolving the white noise field with the real space transfer function, and this choice affects whether the power spectrum or the correlation function is more accurately modeled in a finite box~\citep{Pen1997,Sirko2005,Hahn2011}. Another choice is between grid and glass initial conditions: particle positions at low redshift retain their grid-like structure in voids when initialized on a uniform grid, which some may think has aesthetic drawbacks; on the other hand, the density field from glass initial configurations contains spurious clustering on small scales at high redshift, though this does not grow~\citep{LHuillier2014}. In this paper we investigate the effect of a particular choice -- whether power for $|\mathbf{k}|$ modes greater than the one-dimensional Nyquist frequency, which we call ``corner modes,'' are retained or zeroed.

A Gaussian random field sampled on a lattice can be written as the convolution of white noise with the transfer function~\citep{Efstathiou1985,Salmon1996,Bertschinger2001}:
\begin{equation}
\delta\left(\mathbf{x}\right)=\left(\xi\star T\right)\left(\mathbf{x}\right)=\int d^3x^\prime\,\xi\left(\mathbf{x}^\prime\right)T\left(|\mathbf{x}-\mathbf{x^\prime}|\right),
\label{eqn:transfer}
\end{equation}
where $\xi(\mathbf{x})$ is Gaussian white noise. Setting up initial conditions is thus a problem of representing a random phase realization of a given power spectrum by a discrete set of particles in a simulation volume. This can be carried out by computing the transfer function and white noise field in Fourier space, followed by an inverse Fourier transform, or by convolving an inverse transformation of $T(k)$ with the white noise field as in Equation~\ref{eqn:transfer}. 
Though $P(k)$ is isotropic, because of the Cartesian discretization of $k$-space, the initial conditions are set for $0<$($k_x$,$k_y$,$k_z$)$<$\kny, where \kny$ = \pi N/L$ is the Nyquist frequency, $N$ is the cube root of the total number of particles, and $L$ is the length of the cubic box. This means that there is non-zero power at modes whose moduli are larger than the Nyquist frequency, $|\mathbf{k}|>$\kny, in the corners of the $k$-space cube, i.e., corner modes. 

One can choose to remove corner modes in the initial conditions by applying some additional filtering, such as the spherical Hanning filter used by~\citet{Bertschinger2001}, or by explicitly zeroing the power for $|\mathbf{k}|>$\kny. Note that if the transfer function is sampled in real space, the choice of whether or not to remove corner modes is not explicit, but, e.g., the finite-difference approach of~\citet{Hahn2011} has similar filtering properties to the Hanning filter, with a relatively sharp cutoff in $k$-space. This results in the damping of small-scale power (below \kny), which can be corrected, though these corrections will lead to spectral leakage and associated spurious oscillations~\citep{Hahn2011}. 
Corner modes are often removed in initial conditions generators, but not always, and there is not always a mention of the choice. For example, they are kept in an early version of GRAFIC in order to retain all the power present in the initial density field~\citep{Bertschinger1995}, but they are removed in GRAFIC2 with a note that the effect of the `anisotropy' needs to be studied~\citep{Bertschinger2001}. So far, the effects of the choice of whether to keep or zero corner modes have not been studied in detail, which we aim to do in this paper.

One argument for removing corner modes is that it enforces isotropy and spherical symmetry in Fourier space, but 
we argue here that this is not the case for a discretely sampled density field in a periodic box.

First, suppose we have a continuous sinusoidal signal of $f(x)=\sin(k_0 x)$ in one dimension, which is sampled at regular intervals, $a$. The discrete sampled signal $g(x)$ can be written as
\begin{equation}
g(x) = f(x)  \sum_m \delta(x- m a).
\end{equation}
It turns out that there are infinite frequencies that give exactly the same set of samples. With $k_s = 2\pi/a$, the signal $f_n(x) = \sin((k_0 +n k_s) x) $ sampled the same way gives
\begin{equation}
g_n(x) = f_n(x)  \sum_m \delta(x- m a) = \sin((k_0+n k_s) m a) = g(x).
\end{equation}
This shows that the frequency spectrum contains an infinite series of replicas of the original frequency spectrum shifted by $k_s$, a property known as spectral replication. This periodic replication of the spectrum is entirely due to the discrete sampling.

Now consider a density field sampled on a cubic grid with Nyquist frequency of \kny. For simplicity, let us use an uncorrelated (white noise) random process to populate the density in a box of size $N^3$. The Fourier transform of this field will have independent modes with $N^3$ degrees of freedom, i.e. the corners are filled. As we sample this cubic volume at a regular rate along all three axes, the Fourier space is replicated in a way similar to how periodic boundary conditions behave in configuration space. This spectral replication is a consequence of the axis-parallel, regular sampling.

\begin{figure}
\includegraphics[width=\hsize]{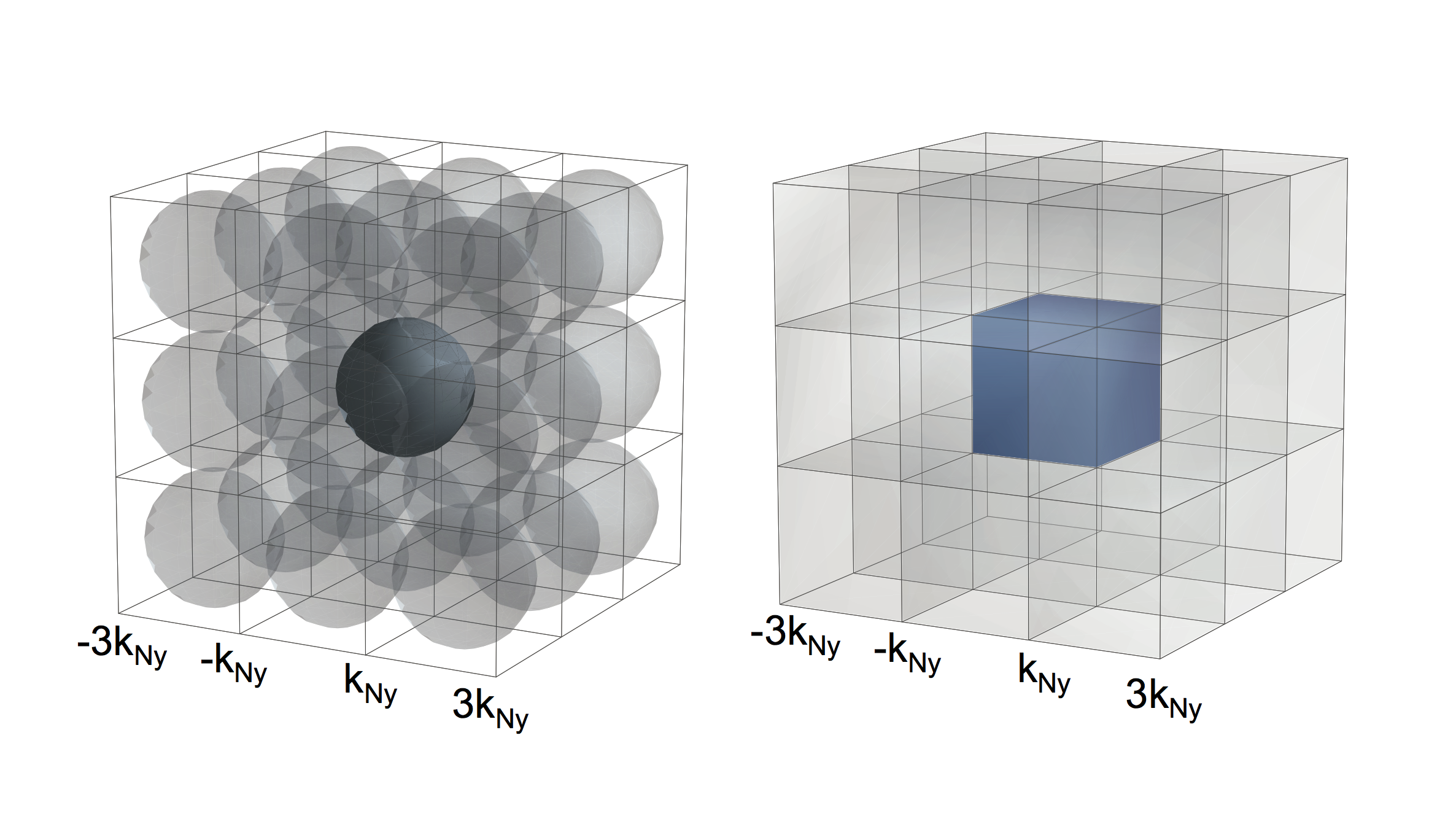}
\caption{We illustrate the effect of the corner mode removal through displaying the envelope of the power for a flat, white noise power spectrum. The grids shown are in Fourier space, and the grids outside the central grid show the power that is mirrored to higher frequencies. From the figure on the left, one can see that the power spectrum after the removal of the corner modes is anisotropic, whereas the figure on the right shows that preserving the corner modes also preserves isotropy.
\label{fig:spheres}}
\end{figure}

This 3-dimensional spectral replication is illustrated in Figure~\ref{fig:spheres}. We show the replicated envelope of the power spectrum in the cases when the corner modes are removed (left) and cases when the corner modes are left intact (right). It is apparent that the replicated pattern is not isotropic when the corner modes are removed, while the Fourier modes fill the available space uniformly when the corner modes are kept. Though the pattern of mirror frequencies occurs on scales smaller than those that were initially resolved, these modes will become relevant at lower redshifts as structure forms on scales smaller than the inter-particle separation. Indeed, as we will show, even if corner modes are removed in the initial conditions, modes at $|\mathbf{k}|>$\kny\ emerge naturally through nonlinear mode-mode coupling and grow as the simulation evolves. Note also that the corner and non-corner modes are entirely independent; there is no aliasing of modes above the Nyquist frequency to lower frequencies since spectral replication only occurs parallel to each axis.

\begin{figure*}
\centering
\includegraphics[width=.9\hsize]{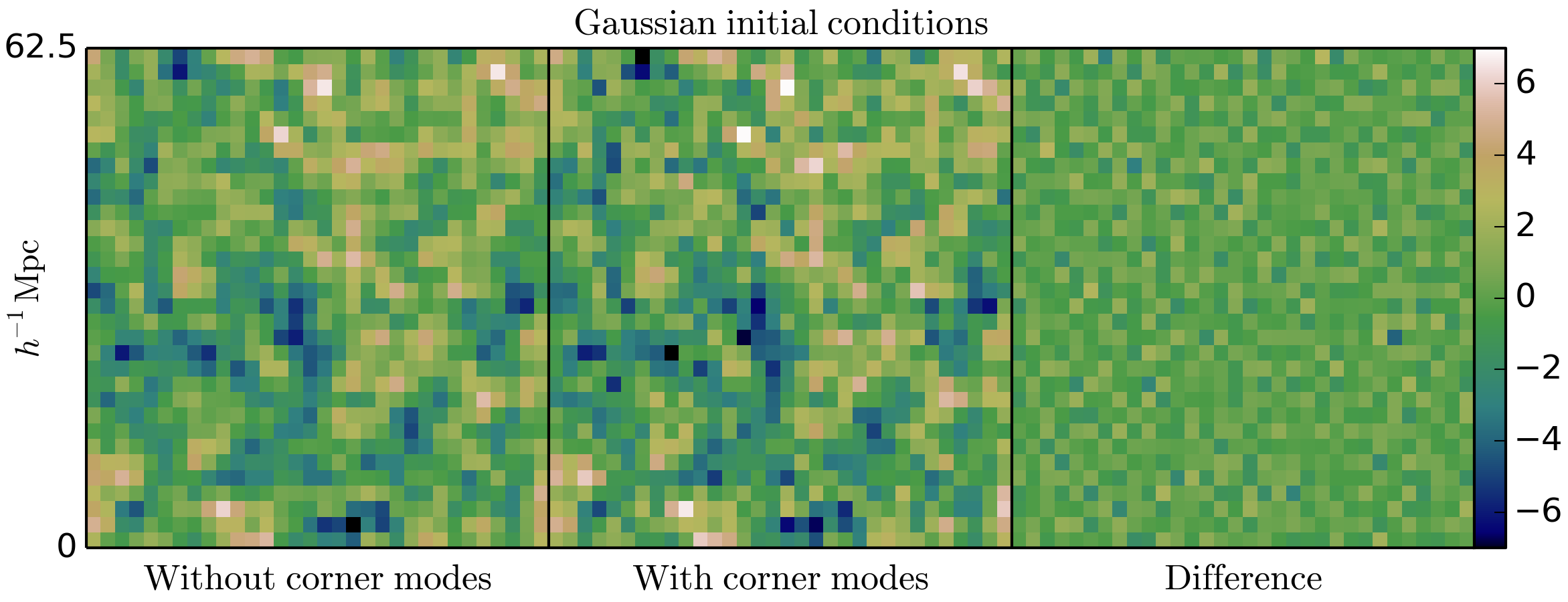}
\caption{Visual differences between initial-conditions density fields with and without corner modes. The density fields were extrapolated in Eulerian linear theory to $z=0$, and the panels are square, 62.5\hmpc\ on a side. These density fields have essentially the properties of the initial conditions of the 512$^3$ particle simulation below, although the realization is different.
\label{fig:cornermodes_difference}}
\end{figure*}

In Figure~\ref{fig:cornermodes_difference} we visually show the difference that corner modes make in the initial conditions for a realization of a Gaussian random field, generated with the same initial power spectrum as used below, linearly extrapolated to $z=0$. Here, we show a 2D, 32$^2$ slice of a 32$^3$ grid of pixels spaced by $\frac{1000}{512}$\hmpc\ (the same spacing as in the 512$^3$ particle simulation used below). 
Zeroing the corner modes visibly smooths the density field; the difference field, essentially a Gaussian field with only corner modes, has visible fluctuations as well.

In this paper we determine the effect of including vs. removing corner modes in the initial conditions of cosmological simulations. In Section~\ref{sec:sims}, we describe the set of simulations used for this test, and in Section~\ref{sec:results} we measure the effect of these corner modes on the matter power spectra, one-point density distribution functions, halo mass functions, and halo density profiles at several redshifts. Conclusions and discussion are presented in Section~\ref{sec:disc}.


\section{Simulations}
\label{sec:sims}

To determine the effect of including or removing corner modes in the initial conditions of simulations, we run two full cosmological dark matter $N$-body simulations where the only difference is whether these modes are present. The simulations were run using the L-Gadget2 code~\citep{Springel2005} with a box length of 1\hgpc, $1024^3$ particles, and WMAP7 cosmological parameters ($\Omega_M = 0.272$, $\Omega_\Lambda = 0.728$, $h = 0.704$, $\sigma_8 = 0.81$, and $n_s = 0.967$)~\citep{Komatsu2011}. Each dark matter particle has a mass of $m_p = 7.031\times 10^{10}$\hmsun, and the Nyquist frequency of the simulation is \kny$ = \pi N/L = 3.2$\ihmpc. The initial conditions are generated using IC\_2lpt\_Gen~\citep{Jenkins2013}, and both simulations have the same random phases. The starting redshift is $z=127$ and was generated using $2^{nd}$ order Lagrangian perturbation theory, sampled on a grid; 64 snapshots are saved down to $z=0$. 
These simulations were run as part of Indra, a suite of 512 simulations with the same box size, particle number, and cosmological parameters with the goal of beating down cosmic variance in large-scale structure studies (Falck, et al., \textit{in preparation}). This means that the box size and resolution are chosen to give good statistics on large scales instead of resolving very small scales.

Even if a difference is found between the simulation where these corner modes are present in the initial conditions and where they are zeroed, it would be unclear which is ``correct.'' To address this, and to disentangle the effects of corner modes and resolution, we also run downgraded versions of both simulations (with and without corner modes) with the same box size of 1\hgpc: one set with only $512^3$ particles, and one set with $1024^3$ particles, but for which the initial conditions were defined up to the Nyquist frequency of a $512^3$ simulation. In this way, we can separately vary only the particle resolution or only the resolution of the initial modes when we do our comparison to determine whether the presence of corner modes has an effect on how well, and at what scales, the lower resolution simulation converges to the `truer' higher resolution simulation. The $512^3$ particle simulations (with and without corner modes) have a dark matter particle mass of $5.62\times 10^{11}$\hmsun\ and a Nyquist frequency of \kny$=1.61$\ihmpc. Unless otherwise specified, results will be presented for the $1024^3$ simulation; at a given $z$ and $k/$\kny, the differences between including and removing corner modes will be smaller/larger for a higher/lower resolution simulation, due to the differences in the onset of nonlinear evolution that washes out the effect of removing corner modes (as we will show).

Halo catalogs are generated as the simulation runs using a standard friends-of-friends algorithm~\citep[FOF;][]{Davis1985}, with a linking length of $b=0.2$ times the inter-particle separation, $L/N$, and a minimum of 20 particles per halo. Post-processing of the FOF halos is performed with SUBFIND~\citep{Springel2001} to identify subhalos in phase space. For the main FOF halo and any subhalos, centers are calculated as the minimum of the potential well, defined by the position of the most bound particle; we use these centers in the following section to measure density profiles of the FOF halos.


\section{Results}
\label{sec:results}

\begin{figure*}
\centering
\includegraphics[width=.9\hsize]{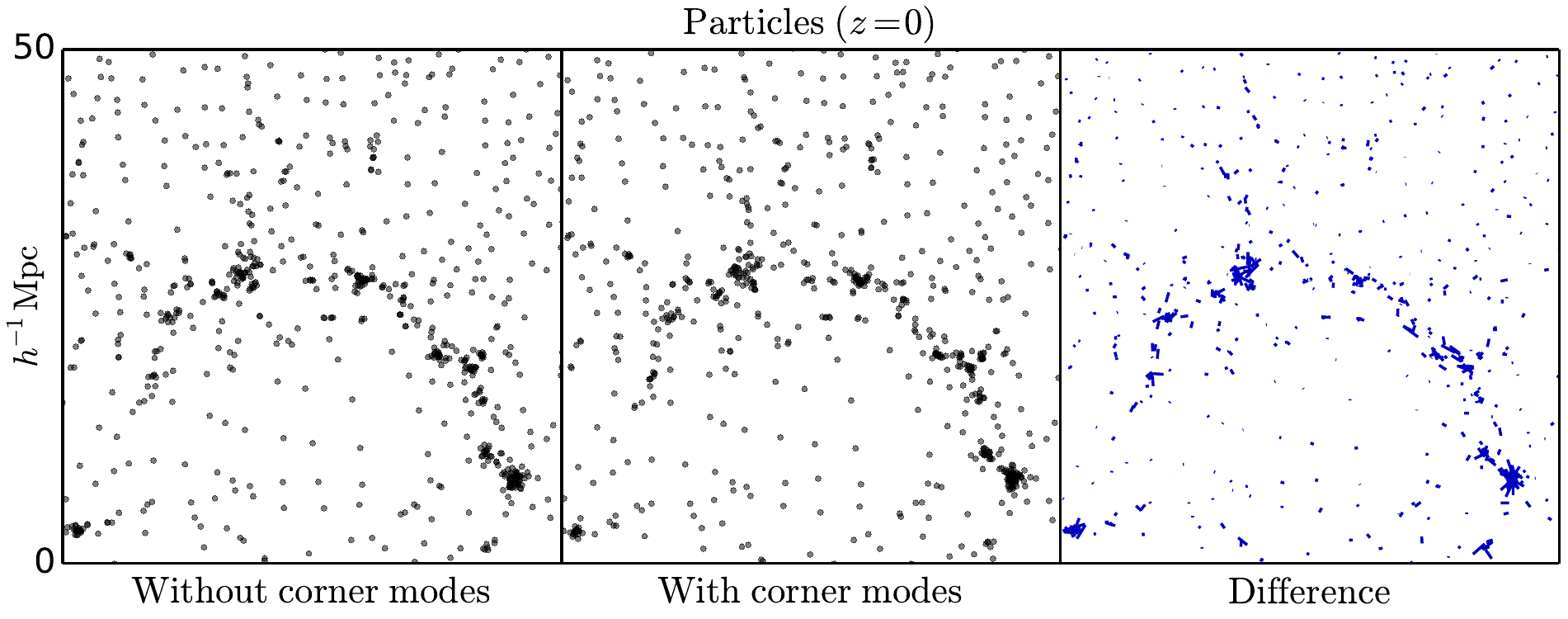}
\caption{Visual differences between $z=0$ particle positions with and without corner modes. The square panels, 50\hmpc\ on a side, show particles on the same 2D initial-conditions sheet, taken from the $512^3$ particle simulations, with initial inter-particle spacing of $\sim 2$\hmpc. On the right, small lines are drawn between positions of particles with and without corner modes.
\label{fig:cmdiffs}}
\end{figure*}

First, in Figure~\ref{fig:cmdiffs}, we visually show the difference in $z=0$ particle positions between the simulations with and without corner modes, from the degraded-resolution $512^3$-particle simulations. The differences are small in voids and larger in high-density regions. At high density, not only is nonlinearity inherently stronger, but there are more close encounters among the particles, giving increased opportunity for numerical noise. Thus, it is not surprising that small initial differences are most amplified in the final conditions at high densities.

In this $512^3$-particle simulation, the distribution of displacements between $z=0$ positions in the simulations with and without corner modes has a median of 0.34 and a root mean square (rms) of 0.61\hmpc. If the displacement between positions with and without corner modes were Gaussian in each dimension, the distribution of squared distances would be a $\chi^2$ with 3 degrees of freedom. Assuming this distribution, an rms distance of 0.61\hmpc\ would give a median distance of $\sim 0.55$\hmpc, much larger than the measurement. This non-Gaussianity in the displacements in each dimension accords with the visual impression of Figure~\ref{fig:cmdiffs} that most displacements are tiny (in low-density areas), with a heavy larger-displacement tail (in high-density regions).

\subsection{Power Spectra}
\label{sec:pk}

In this section, we consider the differences in the power spectra between the simulations with and without corner modes over a range of redshifts. Power spectra were measured using the POWMES code \citep{Colombi2009}, which interpolates the particles to a grid and computes the spectrum using Fast Fourier Transform (FFT). We used a cloud-in-cell (CIC) density assignment scheme on a $1024^3$ grid. POWMES computes the raw power spectrum, as well as the spectrum corrected for the CIC window function and aliasing effects. To study the corner modes in detail, we slightly modified the binning scheme in POWMES to ensure that the particle Nyquist frequency is at the edge of a bin, so that there is no mixing of the corner modes into the bin below the Nyquist frequency.

\begin{figure}
\includegraphics[width=\hsize]{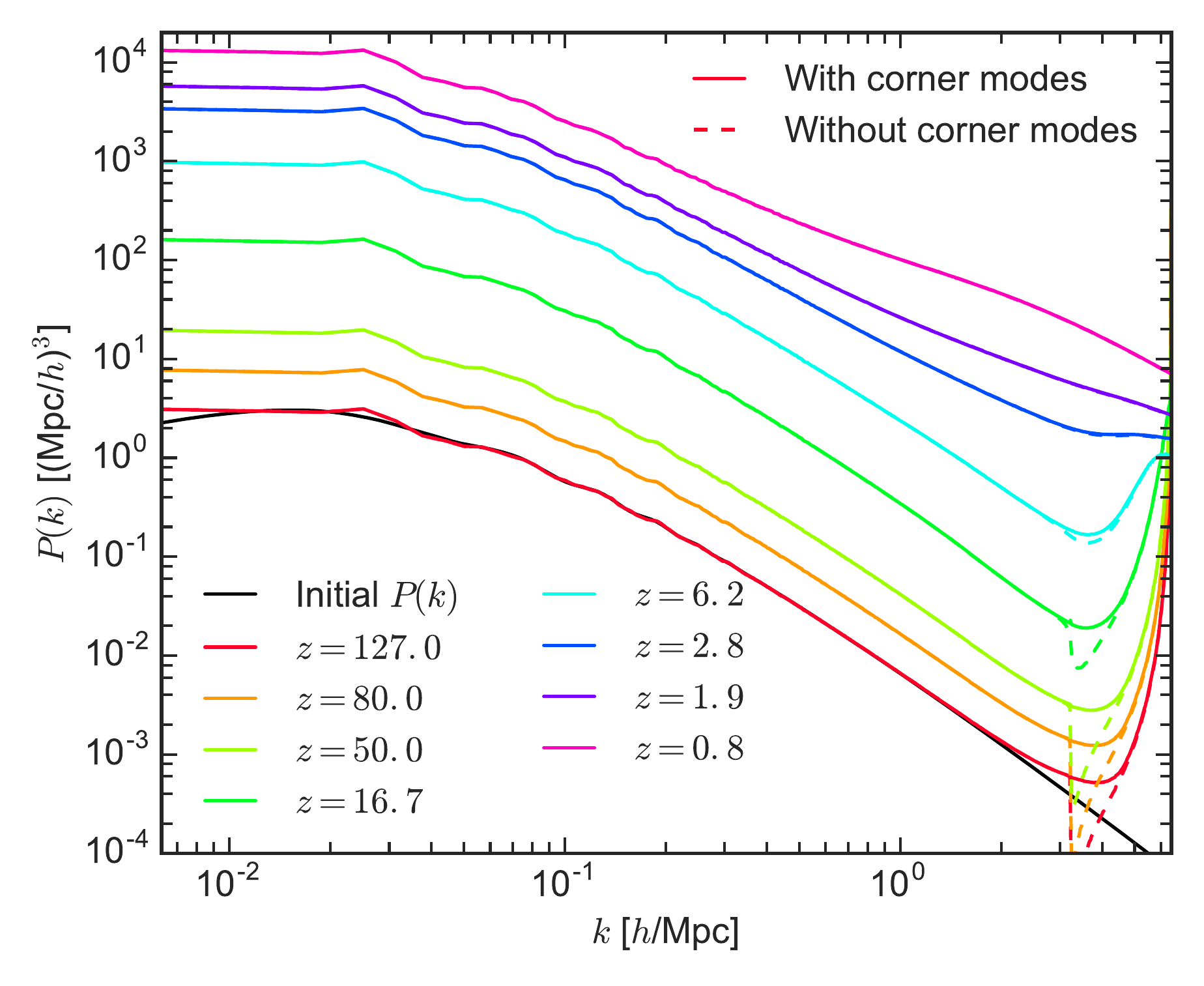}
\caption{Measured power spectra at a range of redshifts covering the evolution of the simulation along with the theoretical input power spectrum. Solid lines show simulations including corner modes and dashed lines show simulations with corner modes zeroed.
\label{fig:pks}}
\end{figure}

Figure~\ref{fig:pks} shows power spectra at several redshifts for both high-resolution ($1024^3$ particles) simulations, along with the initial linear theory power spectrum, computed using CAMB~\citep{Lewis2000}. There is good agreement with the linear theory power spectrum for both simulations at the starting redshift, $z=127$; near the Nyquist frequency (of both the particle distribution and the CIC density grid, \kny$=3.2$\ihmpc), the power spectrum is enhanced above the theoretical power spectrum as expected for a moderately perturbed grid of particles such as is the case for very early snapshots~\citep{Marcos2006,Joyce2007,Colombi2009,Garrison2016}. The simulations with and without corner modes agree very well up to the Nyquist frequency at all redshifts. For $k$ at and just above \kny, at high redshift the power spectrum abruptly cuts off in the simulation where these modes are zeroed. The dramatic upturn in the high-redshift power spectra at wavenumbers above \kny\ is due to the window function correction and is not observed in the raw power spectra. Modes just above \kny\ in the simulation with corner modes removed are slowly restored as $z\to 0$ via nonlinear mode-mode coupling, until both power spectra agree well at all $k$ at around $z=3$. Though we do not explicitly address the growth rate of corner modes in this paper, note that the growth rate of modes is suppressed as one approaches (and exceeds) the one-dimensional Nyquist frequency~\citep{Garrison2016}.

\begin{figure}
\includegraphics[width=\hsize]{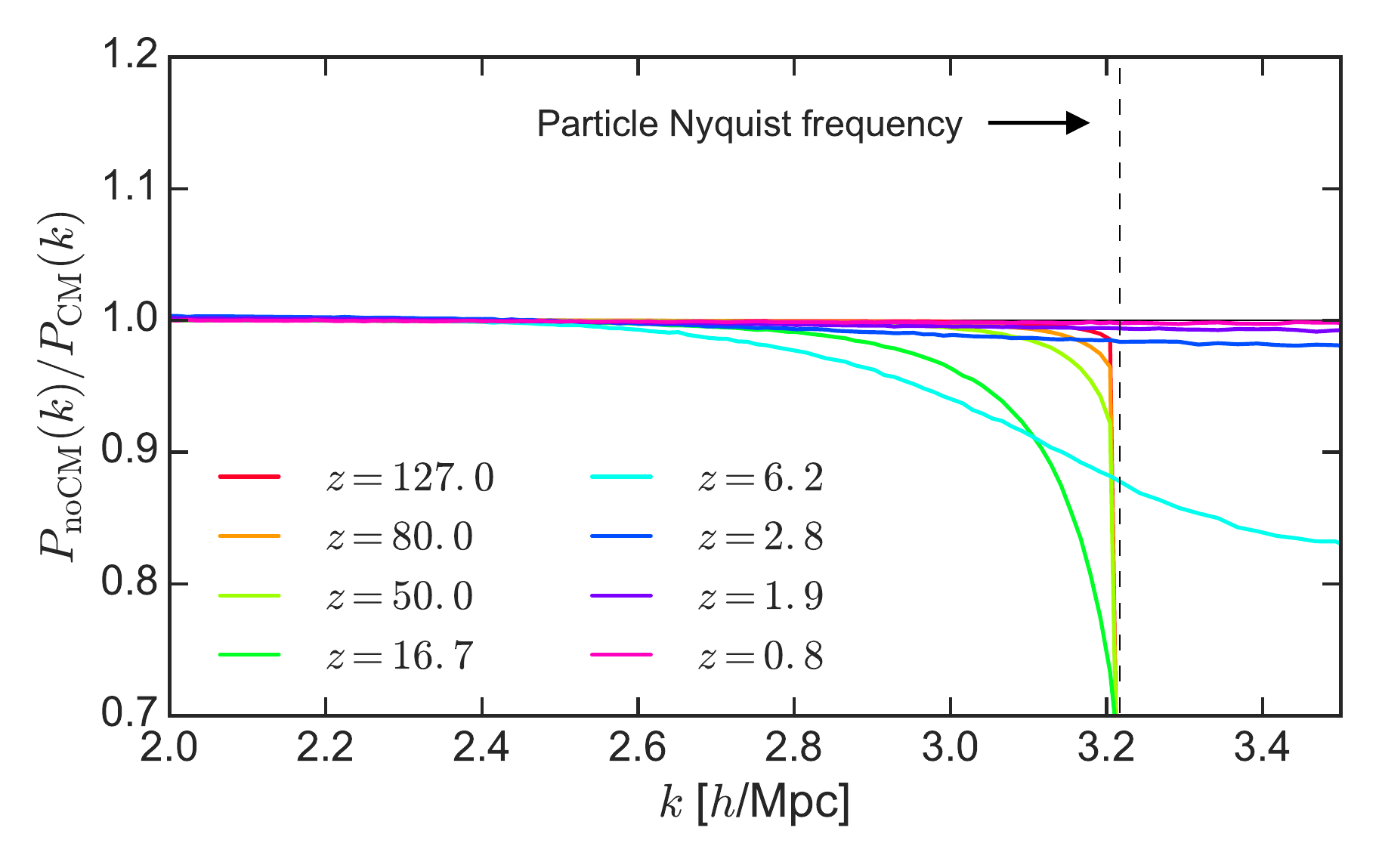}
\caption{Ratio of PS without corner modes to with corner modes at a few redshifts, focusing on the high-$k$ deviation. The vertical dashed line shows the Nyquist frequency, beyond which there is no power in the simulations without corner modes. As redshift decreases, the ratio approaches 1 at all scales.
\label{fig:pkratio_CM}}
\end{figure}

Figure~\ref{fig:pkratio_CM} shows the ratio of power spectra in simulations without corner modes to those where they are included, for the same redshifts as Figure~\ref{fig:pks}. At high-$z$ the ratio is close to 1 up to the Nyquist frequency, where it diverges, as expected. As $z\to 0$ the ratio flattens out, creating differences both above and below \kny, such that the simulation with no corner modes in the initial conditions has less power than the simulation with corner modes at frequencies above about $2.5$ \ihmpc. The deviations in the power spectra below the Nyquist frequency arise from the coupling between the corner modes and lower frequency modes as the simulations progress. By the lowest redshift shown, $z=0.8$, the power spectra from the two simulations very nearly agree, the differences that are present at high redshift having become fully washed out by nonlinear evolution.

\begin{figure}
\includegraphics[width=\hsize]{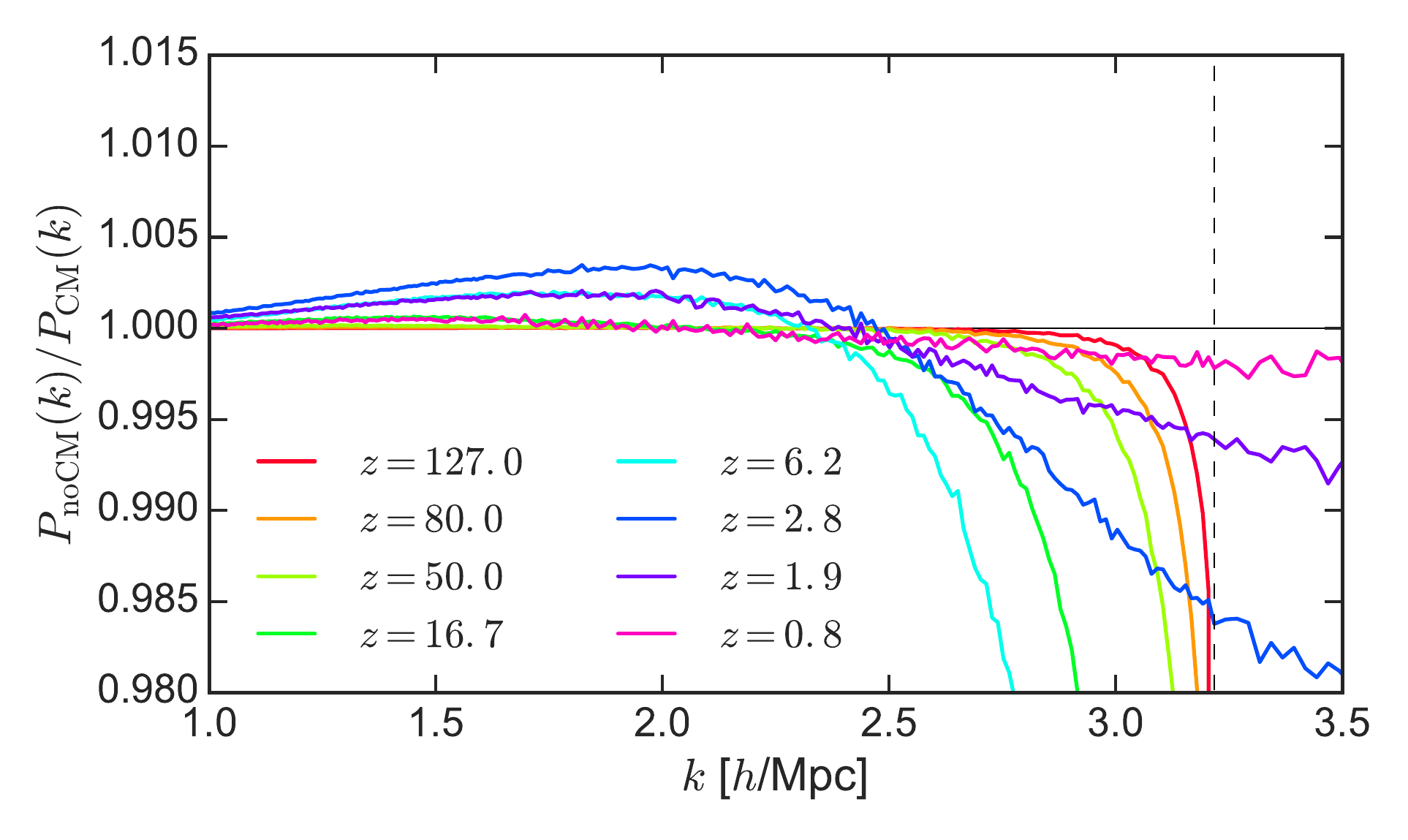}
\caption{The same as Figure~\ref{fig:pkratio_CM}, but focusing on the small percentage deviations at $k>1$\ihmpc.
\label{fig:pkratio_CMzoom}}
\end{figure}

We examine these differences in greater detail in Figure~\ref{fig:pkratio_CMzoom}, which is the same as Figure~\ref{fig:pkratio_CM} but is zoomed-in to show small percentage deviations at $k>1$ \ihmpc. Interestingly, at intermediate redshifts, there is an increase in power in the simulation with no corner modes relative to the simulation with corner modes for frequencies $k<2.5$ \ihmpc. The maximum deviation is $\sim 0.4\%$ at $k\sim 2$\ihmpc\ at $z=2.8$. The deviation decreases below $z=2.8$, and the ratio is again close to 1 for $z=0.8$. Note that for the lower particle resolution simulations, these differences persist to lower redshifts due to the delayed repopulation of the corner modes.

\begin{figure}
\includegraphics[width=\hsize]{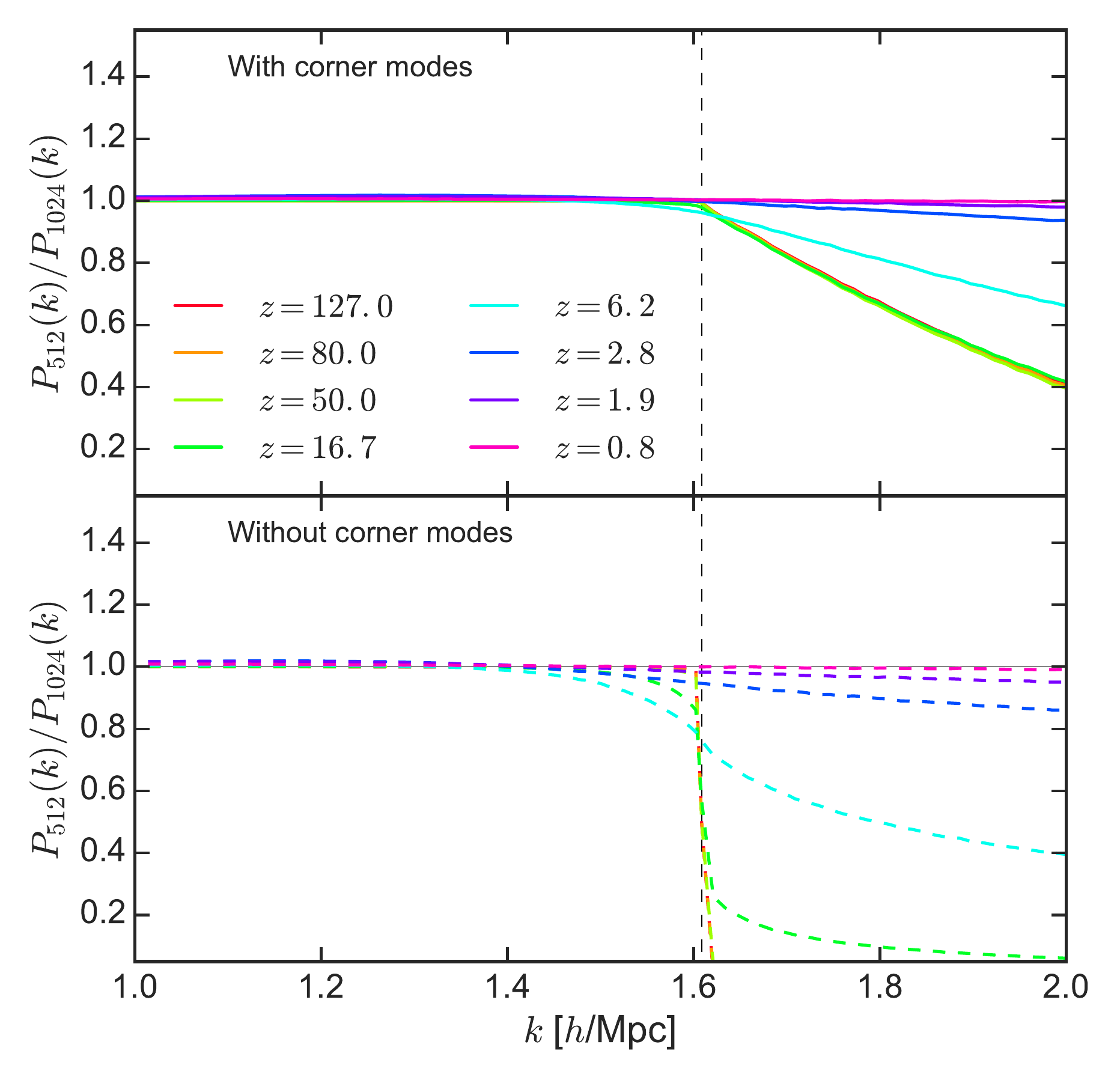}
\caption{Ratio of the PS of the high-resolution simulation with downgraded initial modes to the PS of the high-resolution simulation with all initial modes, for simulations with (solid lines, upper panel) and without (dashed lines, lower panel) corner modes.
\label{fig:pkratio_vary_modes}}
\end{figure}

\begin{figure}
\includegraphics[width=\hsize]{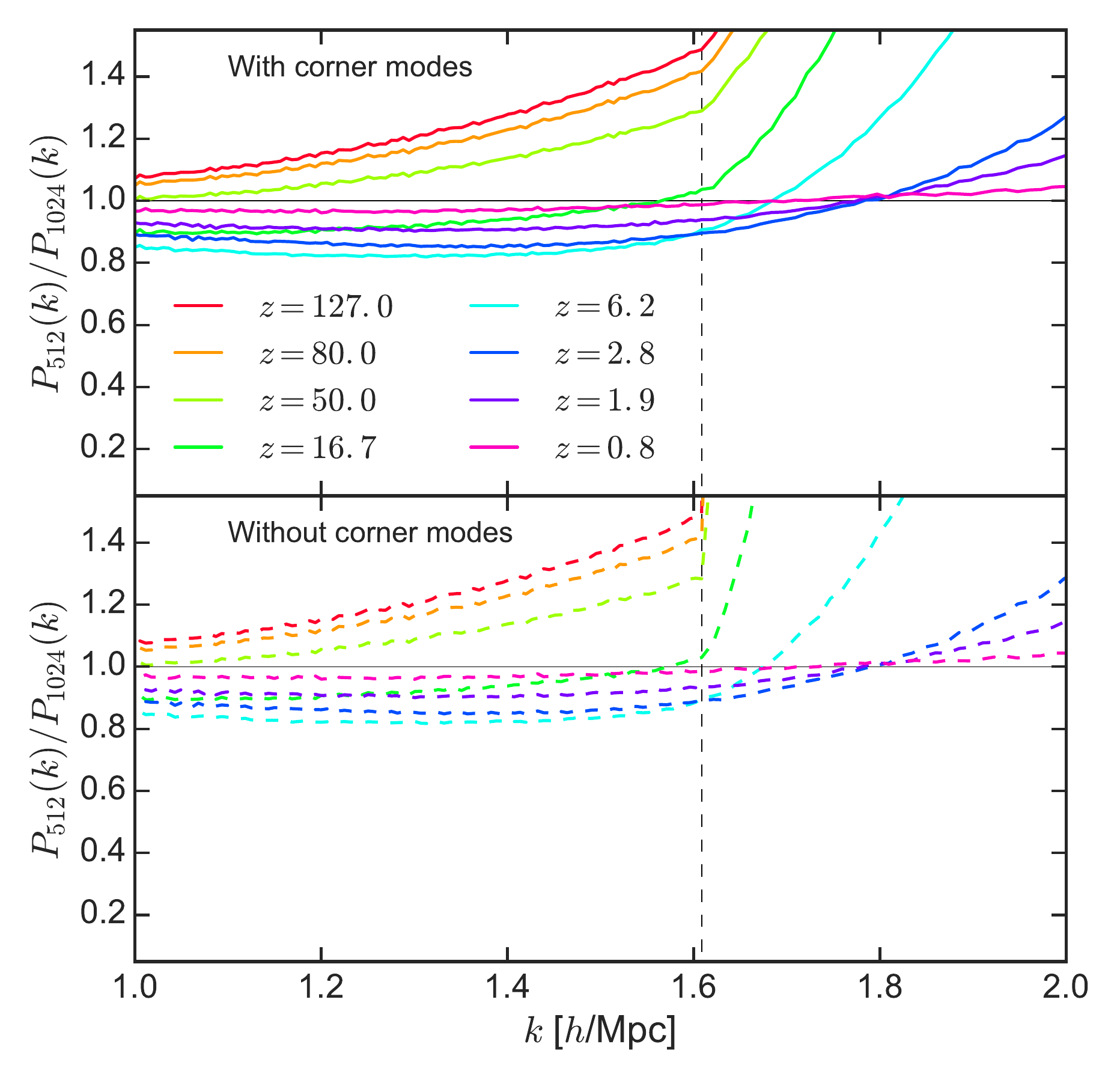}
\caption{Ratio of the PS of the $512^3$ particle simulation to the PS of the $1024^3$ particle simulation with $512^3$ initial modes, for simulations with (solid lines, upper panel) and without (dashed lines, lower panel) corner modes. 
\label{fig:pkratio_vary_res}}
\end{figure}

In order to determine whether including corner modes is ``better'' than zeroing them out in the initial conditions, we make two comparisons: in Figure~\ref{fig:pkratio_vary_modes} we plot power spectrum ratios of the $1024^3$ particle simulations, where the `512' simulation has had initial modes defined up to the Nyquist frequency of a $512^3$ particle simulation, such that we are only varying the resolution of the initial modes; and in Figure~\ref{fig:pkratio_vary_res} we plot power spectrum ratios of standard $512^3$ particle simulations to these downgraded simulations with $1024^3$ particles, such that we are only varying the resolution of the dark matter particles. 
In both figures, the top panel shows the ratio for the simulations with corner modes, and the bottom panel shows the ratio for simulations without corner modes. In Figure~\ref{fig:pkratio_vary_modes}, the ratios are near unity at low-$k$ until the Nyquist frequency of the simulation with the downgraded initial modes (given by the vertical dashed line at $k=1.6$ \ihmpc). Above this frequency, the ratio drops drastically at high redshifts when corner modes are removed, and it drops gradually when they are retained. Both ratios return to unity as the redshift decreases, remaining closer to unity in the case where corner modes are retained. This suggests that while retaining corner modes above the one-dimensional Nyquist frequency does not result in an `exact' power spectrum at these frequencies, it is better than imposing a sharp-$k$ cutoff, at least at high redshifts.

Now we turn to the question of particle resolution. In Figure~\ref{fig:pkratio_vary_res}, for the earliest snapshots, both ratios are enhanced above unity near the Nyquist frequency of the particles in the lower-resolution simulation (given by the vertical dashed line at $k=1.6$ \ihmpc). This is because the particles are on a very nearly uniform grid, as we discussed for the high-redshift power spectra in Figure~\ref{fig:pks}. The ratios in both cases, with and without corner modes, show a sharp upturn at this frequency at high redshifts, and are slightly sharper in the case without corner modes. The ratios dip below unity and then again approach unity as redshift decreases, and the discontinuity smooths out, in both cases.

These results suggest that including corner modes in a lower mode-resolution simulation results in a power spectrum that is closer to its higher mode-resolution counterpart. However, these effects are much smaller than particle resolution effects, which cause the power spectra in both cases to differ from the high resolution case by as much as 20\%, even after the particles are sufficiently displaced from the grid. Also, the particle resolution affects a much wider range of wavenumbers ($0.6 < k <$ \kny \ihmpc) than the corner modes, which only affect the power spectra very close to and above the one-dimensional Nyquist frequency. 

\subsection{One-point Probability Distribution Functions of Density}

In this section we explore the effect of the corner modes on the one-point probability distribution functions (PDFs) of the overdensity at various redshifts. The PDF is the lowest-order statistic of the matter-density field. It is sensitive to phase correlations and therefore visual differences in the density field~\citep{Neyrinck2014}. The PDF can be difficult to model, but recently, spherical dynamics have been shown to be remarkably accurate to model it, both for a mass-weighted PDF~\citep{Neyrinck2013} and a volume-weighted PDF~\citep{BernardeauEtal2013,UhlemannEtal2016}. A matter-density PDF, if it can be inferred observationally, would provide impressive cosmological constraints~\citep{CodisEtal2016}. 

\begin{figure}
\includegraphics[width=\hsize]{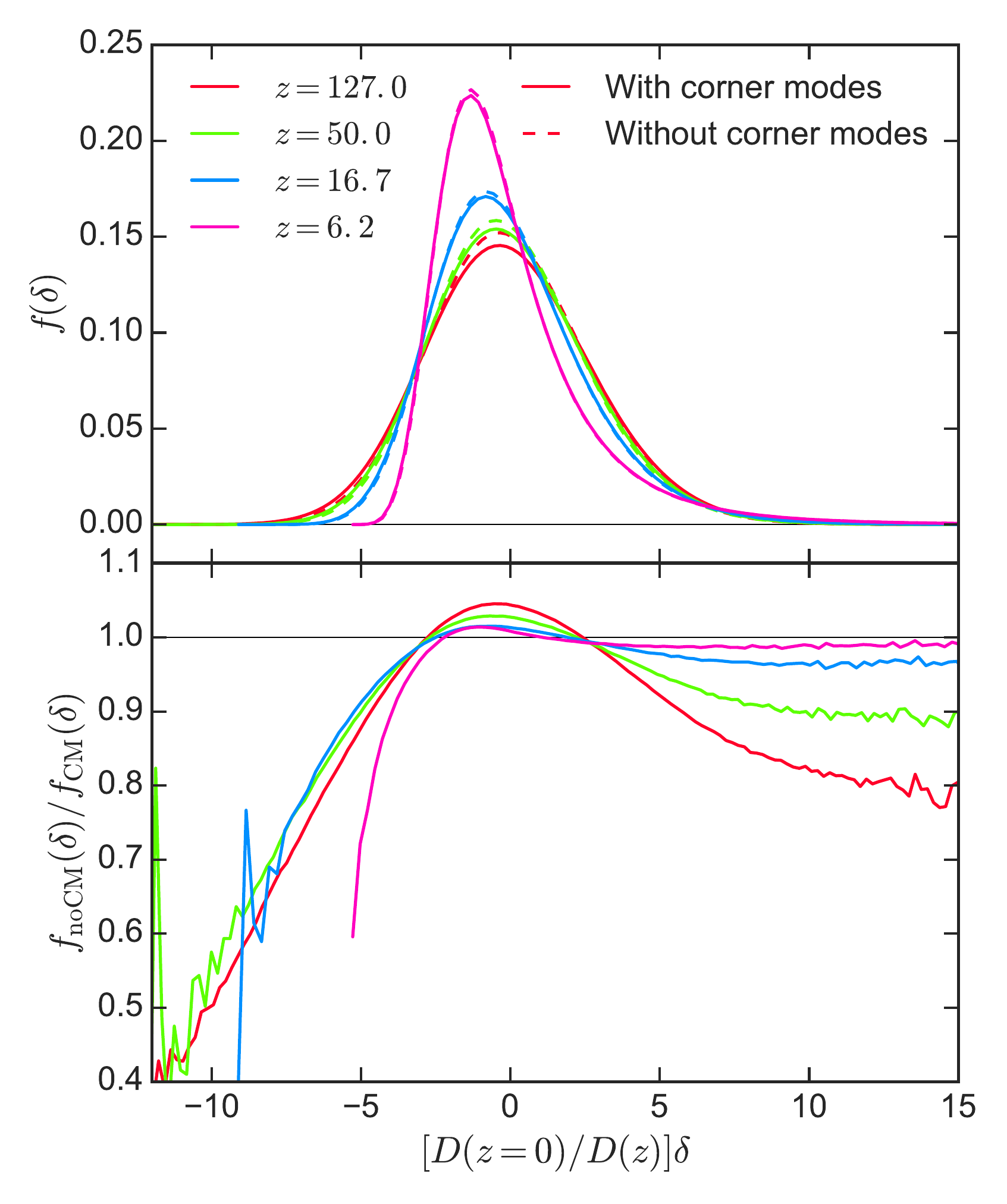}
\caption{Top: one-point distribution functions of the overdensity field using CIC density assignment in 1 \hmpc\ cells from simulations with and without corner modes, at several redshifts. Bottom: ratio of distribution functions without corner modes to those with corner modes.
\label{fig:pdf}}
\end{figure}

We compute the overdensity field using a CIC assignment scheme with $\frac{1000}{1024}\approx1$\hmpc\ cells. The overdensity PDFs are plotted in Figure~\ref{fig:pdf} at several redshifts. To aid in the comparison, we scale $\delta$ with the inverse growth factor, [$D(z=0)/D(z)$], allowing us to compare just the shapes of the distributions. We only show PDFs for $z>6.2$ because the differences are negligible at lower redshifts.

The bottom panel of Figure~\ref{fig:pdf} shows the ratio of the PDFs in the simulation without corner modes to that with corner modes. The PDFs without corner modes are narrower, thus the ratio is suppressed at low and high densities and enhanced near $\delta=0$. There is a strong redshift dependence at high densities: removing corner modes suppresses high densities by 20\% at the initial redshift, reducing to $\sim 5\%$ at $z=16.7$ and $\sim 1\%$ at $z=6.2$ as these modes become repopulated during the evolution of the simulation.

\begin{figure}
\includegraphics[width=\hsize]{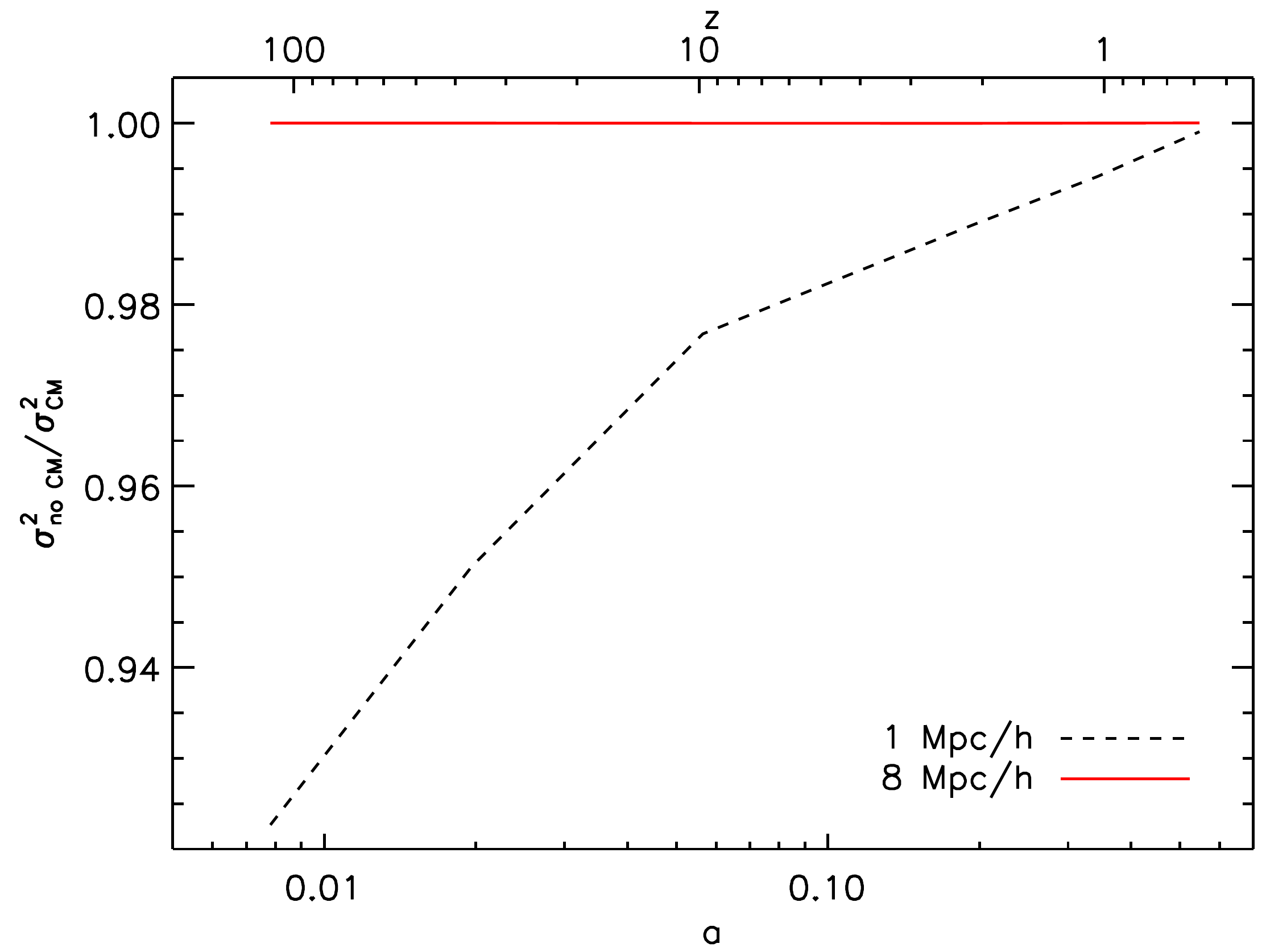}
\caption{Ratio of the variance of the one-point PDF without corner modes to that with corner modes as a function of redshift, using both 1\hmpc\ cells (dashed line) and 8\hmpc\ cells (solid line).
\label{fig:variance}}
\end{figure}

To quantify the evolution of the difference in the density distribution functions with redshift, we measure the variance of the PDFs using both 1\hmpc\ and 8\hmpc\ CIC grid cells at several redshifts. The ratios of these variances in the simulations without corner modes to those with corner modes are shown in Figure~\ref{fig:variance}. Zeroing the corner modes reduces the 1\hmpc\ variance in the density field at high redshift compared to leaving them, and this difference reduces to $\sim 0.05$\% at $z=1$, while the difference in the 8\hmpc\ variance is negligible.

Note that this variance ratio depends on the logarithmic slope of the initial power spectrum at \kny\ and therefore also on the simulation resolution. If the slope is steep (at high resolution, for a typical $\Lambda$CDM simulation), the fractional power removed at the initial inter-particle scale by zeroing corner modes will be less than that for a shallower slope. 
Assuming the same $\Lambda$CDM linear power spectrum as that used in these simulations, changing the simulation resolution by a factor of 10 changes the effect of corner modes on the variance of the overdensity PDF by a few percent.

\subsection{Halo Mass Functions}

\begin{figure}
\includegraphics[angle=-90,width=\hsize]{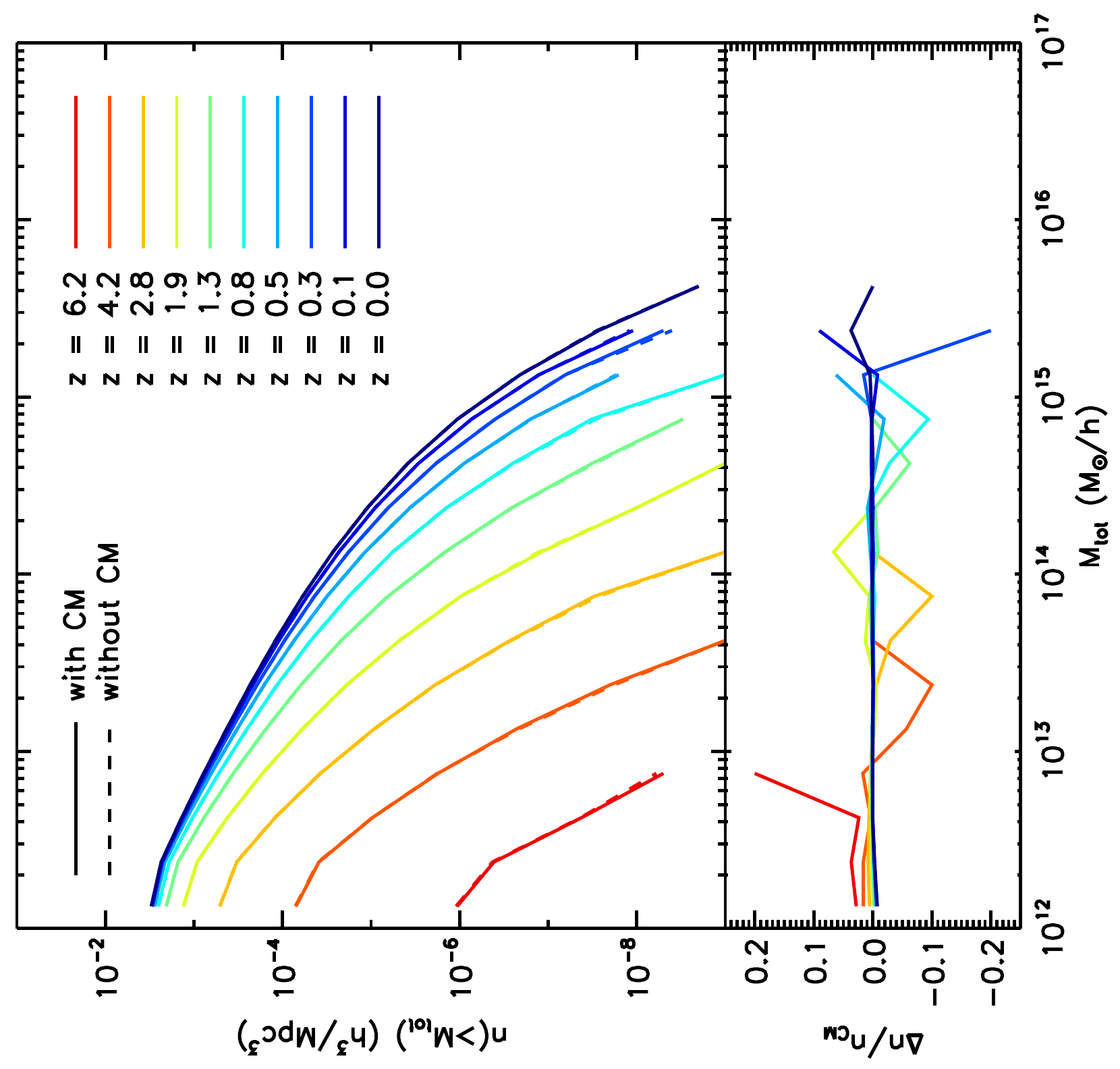}
\caption{Top panel: total mass functions of FOF halos from $z=6.2$ to $z=0$ for the simulation with (solid lines) and without (dashed lines) corner modes. Bottom panel: ratio of mass functions without corner modes to those with them.
\label{fig:massfn}}
\end{figure}

In Figure~\ref{fig:massfn} we plot halo total mass functions at 10 redshifts from $z=6.2$ to $z=0$. Note that these are not mass functions for virial masses or $M_{200}$, since we did not want our results to depend on additional post-processing of the halos. There are tiny differences in the cumulative mass functions of the simulations with and without corner modes in the initial conditions, mostly at the high-mass end. Though these differences occur at all redshifts, they are not systematic -- sometimes mass functions from the simulation with corner modes are higher than those without corner modes, sometimes they are lower -- and the differences are very small. This can be seen in the bottom panel of Figure~\ref{fig:massfn}, which shows the ratio of the mass functions from the simulation without corner modes to the simulation with corner modes. The largest differences, up to 20\% but mostly within 10\%, are in the largest mass bins at a given redshift, where statistics are poor. 

There is also a small difference in the number of small halos with total masses less than $3\times 10^{12}$\hmsun. Note that the smallest halos, containing 20 particles, have masses of $1.3\times 10^{12}$\hmsun. At $z=6.2$, there are 3\% more small halos in the simulation without corner modes; this difference reduces as $z\to 0$ and is less than 1\% by $z=2.8$. Interestingly, the ratio dips slightly below zero at $z=2$ and continues to decrease, such that by $z=0$ there are 0.7\% fewer small halos in the simulation without corner modes than with corner modes. Recall that the ratio of power spectra without corner modes to those with corner modes is also increased between $k= 1$ and $k = 2.5$\ihmpc\ until $z\sim 2$ (see Figure~\ref{fig:pkratio_CMzoom}), so this seems to be also reflected in the low end of the mass functions.

\begin{figure*}
\centering
\parbox{0.45\textwidth}{\includegraphics[angle=-90,width=\hsize]{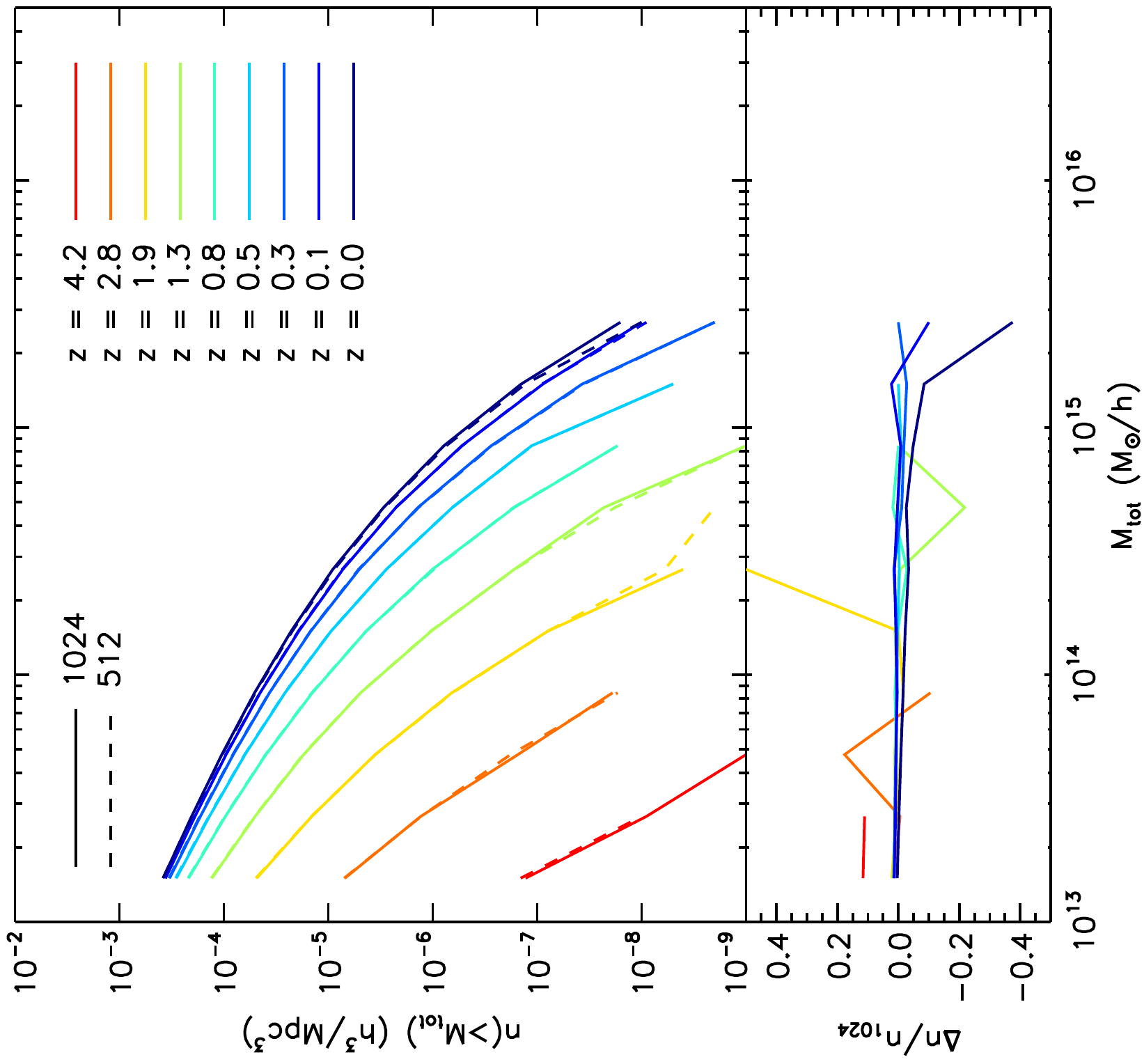}}%
\qquad
\begin{minipage}{0.45\textwidth}%
\includegraphics[angle=-90,width=\hsize]{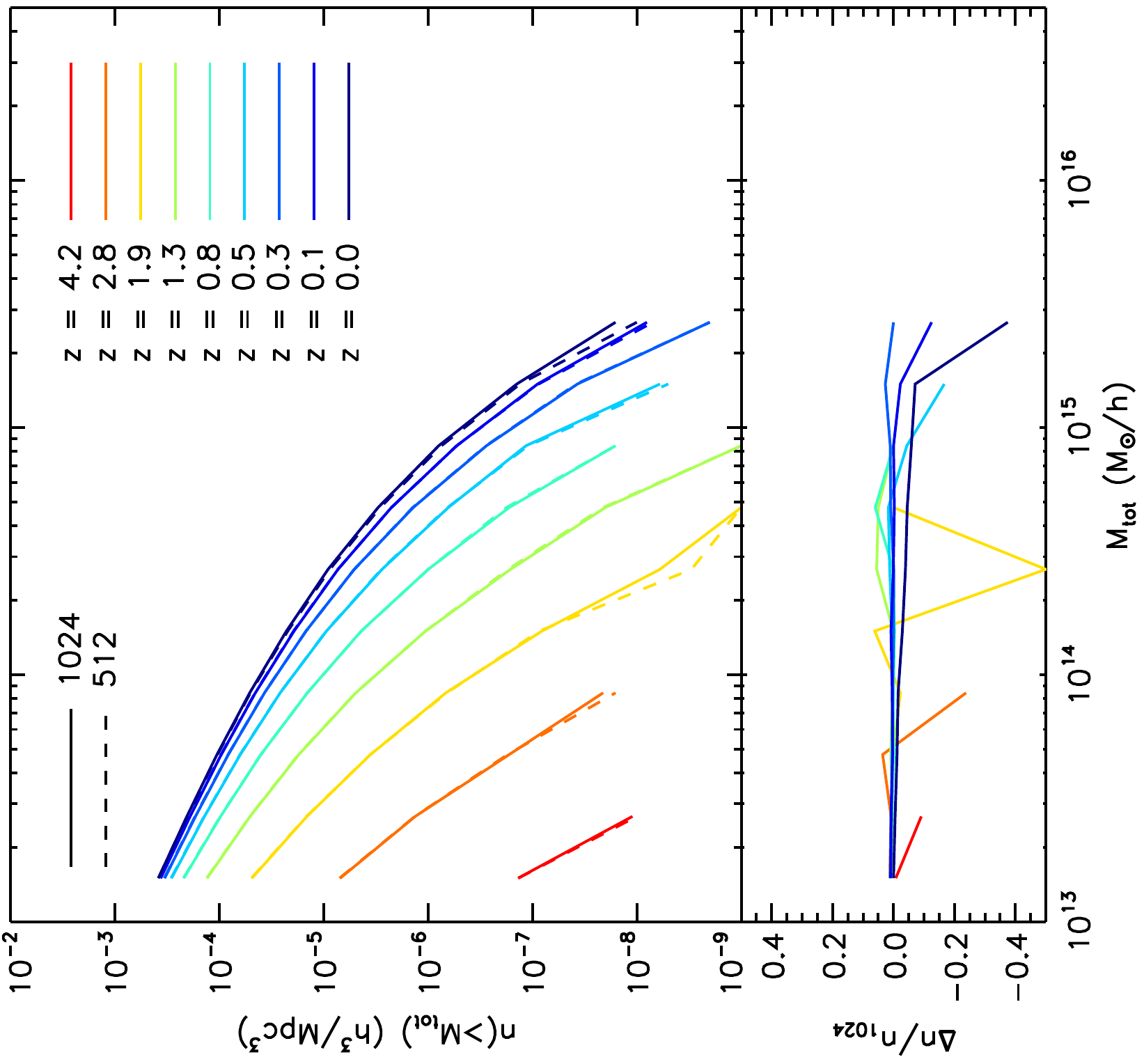}
\end{minipage}%
\caption{Total mass functions of FOF halos from $z=4.2$ to $z=0$ for the simulations without corner modes (left) and the simulation with corner modes (right), with $1024^3$ particles and full initial $k$-modes (solid lines) and with low-resolution initial $k$-modes defined on a $512^3$ grid (dashed lines). All simulations have $1024^3$ particles in a 1\hgpc\ box. Bottom panels: ratio of mass functions with low-resolution initial $k$-modes to those with full-resolution initial $k$-modes.
\label{fig:massfn_res}}
\end{figure*}

Given these differences in the low end of the mass functions, we would like to have some idea of whether including or removing corner modes is ``better'' independent of the particle resolution, as we did with Figure~\ref{fig:pkratio_vary_modes}. In Figure~\ref{fig:massfn_res} we plot the mass functions of both $1024^3$ particle simulations, where the ``512'' simulation was initialized with modes only up to the Nyquist frequency of a $512^3$ particle simulation, as described in Section~\ref{sec:sims}. Due to the suppressed small-scale initial power of the `512' simulation, there are fewer halos at high redshift, so we start at $z=4$. The mass functions without (left) and with (right) corner modes are shown in the upper panels, and the bottom panels show the ratios of the lower to higher resolution mass functions. The mass functions and ratios are qualitatively similar for both simulations with and without corner modes, and the largest differences are in the highest-mass bins where there are only a few halos. 

It is interesting to note how the lack of initial power in the ``512'' simulations leads to a suppression of the high end of the mass functions at low redshifts, as seen in the bottom panels; however, this behavior is similar in both simulations with and without corner modes, so we find that removing corner modes has no significant effect on the mass functions of dark matter halos. This can be understood if the onset of nonlinear structure formation washes out the effect of removing corner modes by repopulating them, which happens at the same redshift as halo formation, regardless of simulation resolution.

\subsection{Halo Profiles}

As seen in Figure~\ref{fig:pdf}, the largest difference in the density one-point PDF is at high densities, thus it may be that the presence or absence of corner modes has some effect on the inner regions of halos. To test this, we measure stacked profiles of all halos (in different mass bins) at several redshifts. Halo centers are measured in SUBFIND for the FOF parent halos as the position of the most bound particle, i.e., the center of the potential well. The density profiles of each halo are measured using all halo particles and then stacked in logarithmic radial bins. Error bars show the standard deviation of the mean halo profile in each radial bin. Though all halos have at least 20 particles, we stack profiles in two bins of particle number (and therefore total mass): 30 to 100 particles (or $2.11\times 10^{12}$ to $7.03\times 10^{12}$\hmsun), and greater than 100 particles. The minimum radius of the stacked profile is set to the largest minimum radius of the individual profiles of both simulations.

\begin{figure}
\includegraphics[angle=-90,width=\hsize]{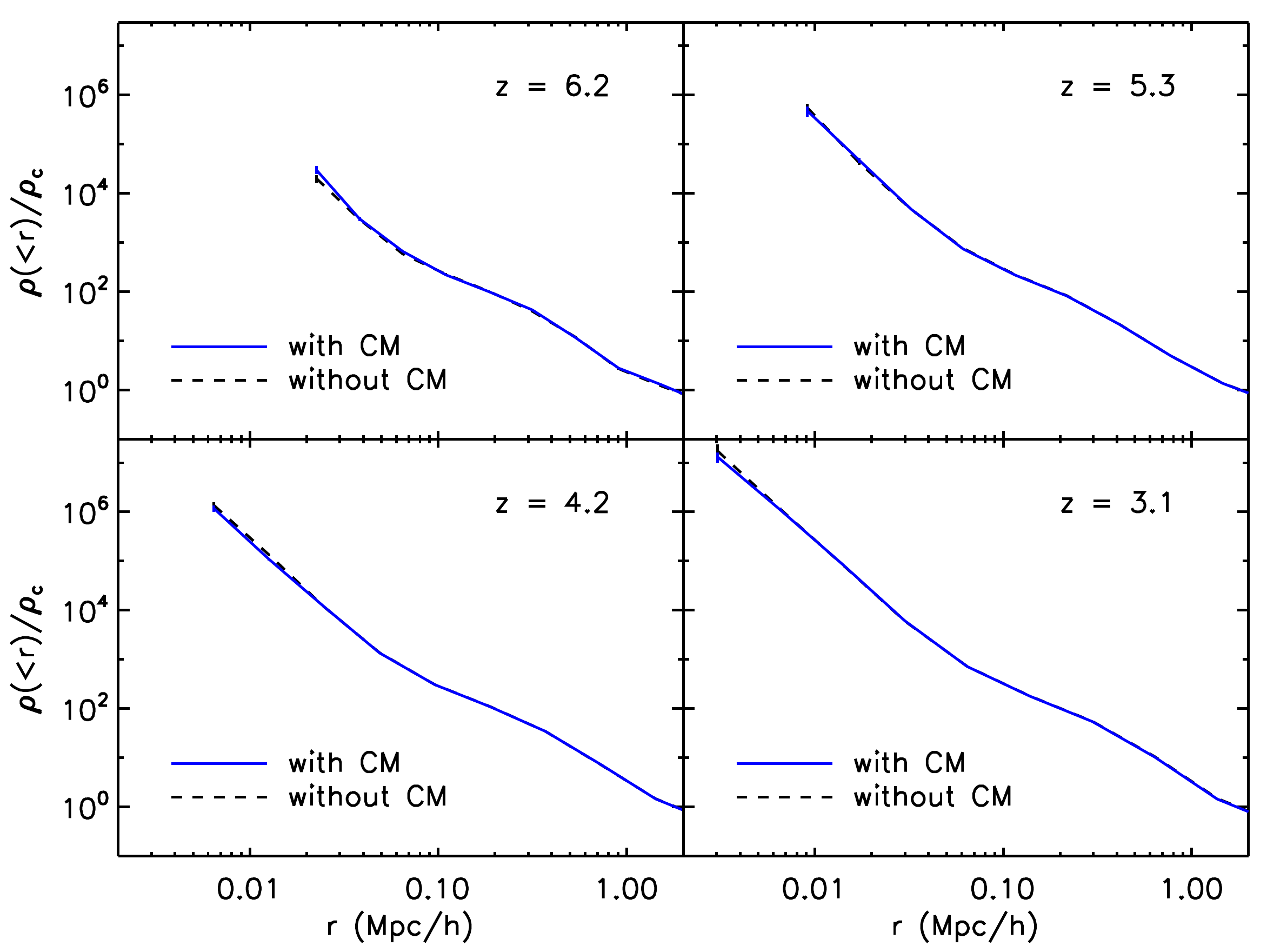}
\caption{Halo density profiles at $z=6.2$, 5.3, 4.2, and 3.1 for halos with greater than 30 particles and fewer than 100 particles.
\label{fig:profs_small}}
\end{figure}

\begin{figure}
\includegraphics[angle=-90,width=\hsize]{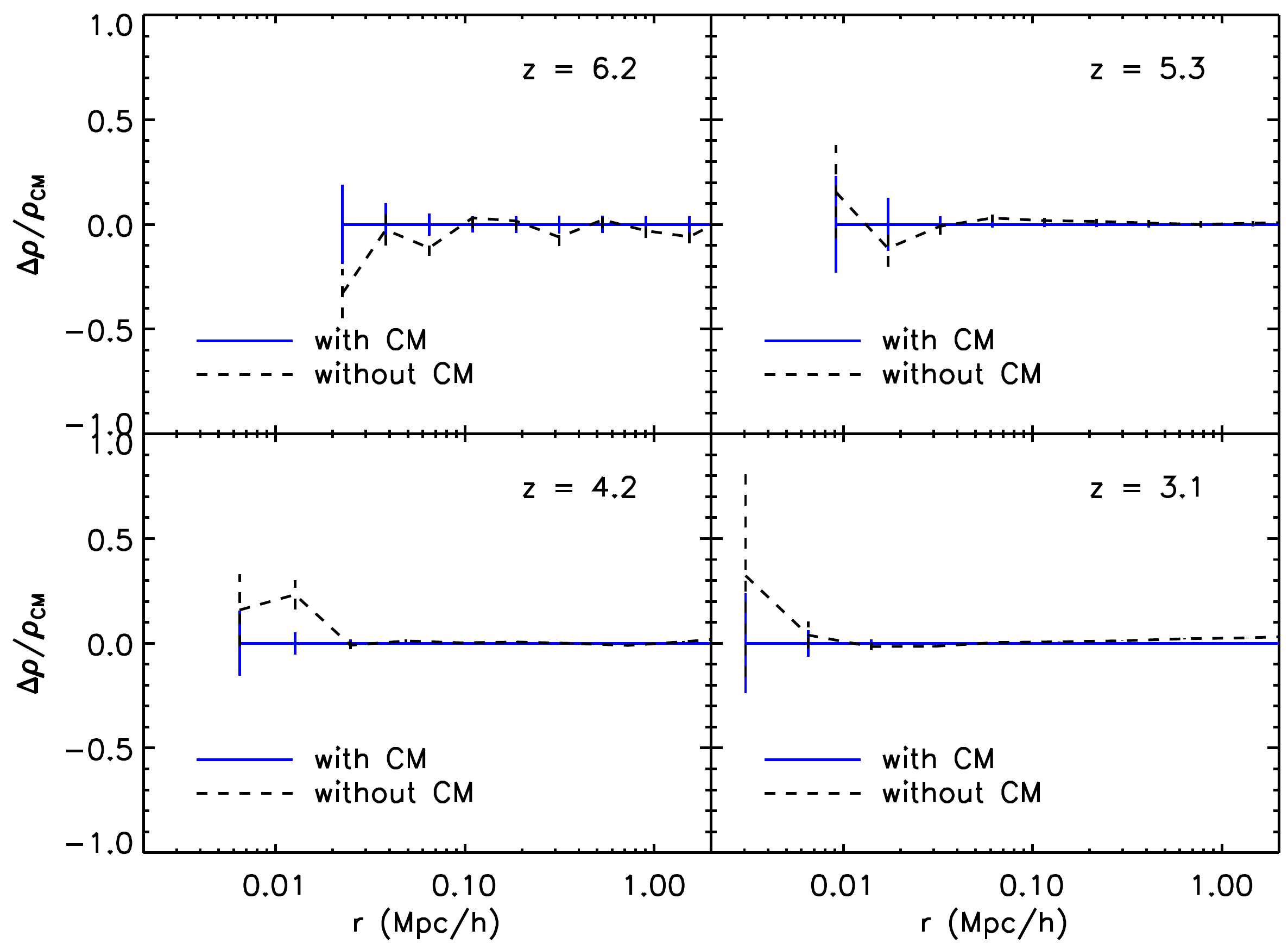}
\caption{Ratio of stacked halo density profiles of the simulation without corner modes to that with corner modes, at $z=6.2$, 5.3, 4.2, and 3.1 for halos with greater than 30 particles and fewer than 100 particles.
\label{fig:profs_small_ratio}}
\end{figure}

Figure~\ref{fig:profs_small} shows the stacked density profiles of halos with 30 to 100 particles at 4 redshifts from $z=6.2$ to $z=3.1$. There is very little difference between stacked profiles from simulations with and without corner modes in the initial conditions. As the redshift decreases, there are more halos and the profiles probe smaller radii. The ratios of the profiles from the simulation without corner modes to that with them are shown in Figure~\ref{fig:profs_small_ratio}, for the same profiles as in Figure~\ref{fig:profs_small}. The differences are clearly greater for the inner regions of the profiles, but this is where the profiles are not well resolved, and the differences are within the scatter given by the error bars. Note that at redshifts higher than $z=6.2$, most of the halos have only 20 particles, and at redshifts lower than $z=3.1$ the differences are even smaller.

\begin{figure}
\centering
\includegraphics[angle=-90,width=\hsize]{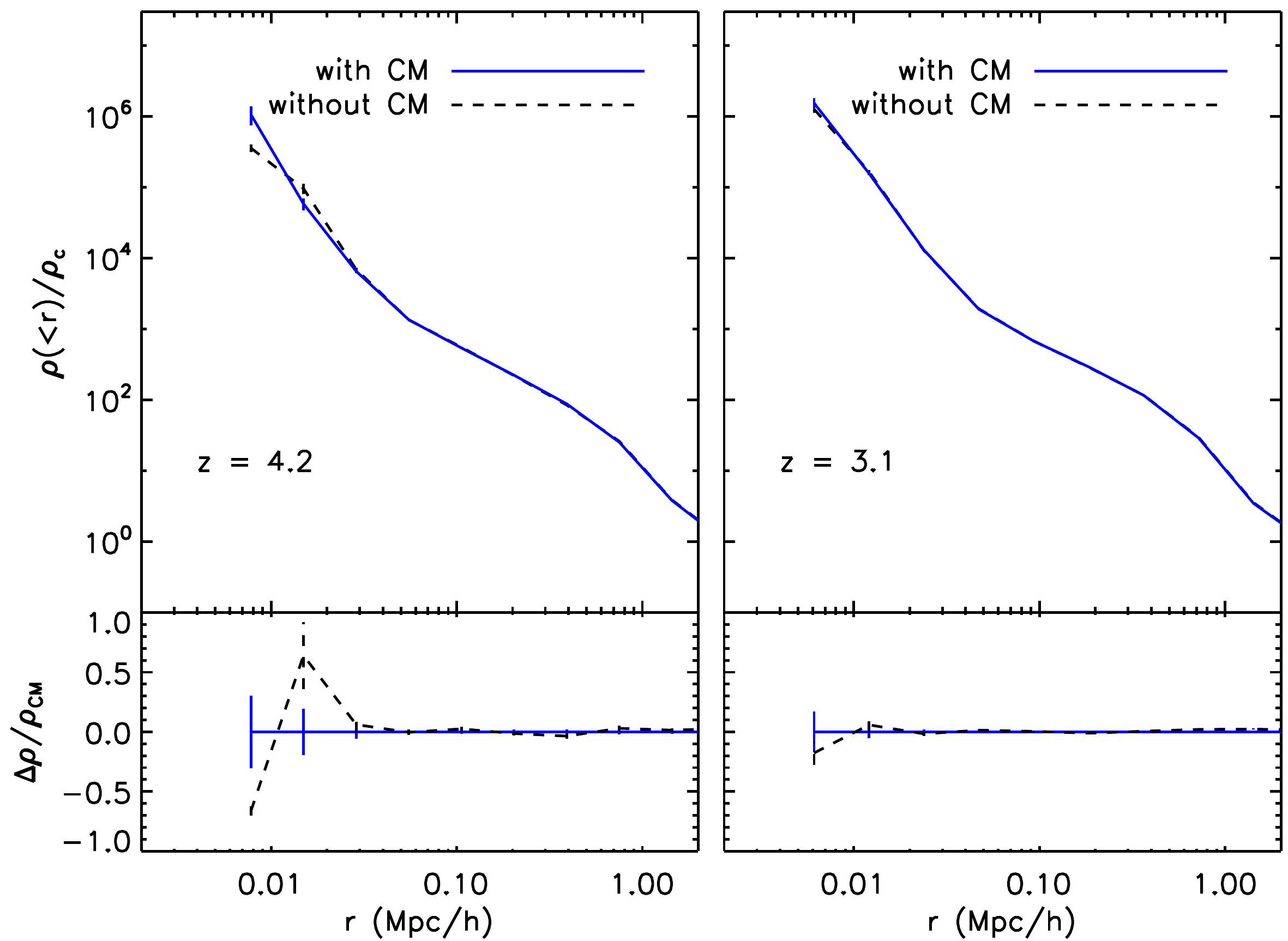}
\caption{(Top) Stacked halo density profiles and (bottom) ratio of halo density profiles in the simulation with corner modes to those in the simulation without corner modes, at $z=4.2$ (left) and $z=3.1$ (right) for halos with greater than 100 particles.
\label{fig:profs_large}}
\end{figure}

At $z=6.2$ and $z=5.3$, there are very few halos with more than 100 particles, so in Figure~\ref{fig:profs_large} we show the stacked profiles at $z=4.2$ and $z=3.1$, respectively, for halos with greater than 100 particles ($M_{\rm{tot}}> 7.031\times 10^{12}$\hmsun). In the bottom panels we plot the ratio of profiles from the simulation without corner modes to that with corner modes. At both redshifts, the profile from the simulation with the corner modes included is enhanced with respect to the profile with no corner modes in the smallest radial bin. This difference appears to be more significant for $z=4.2$ than for $z=3.1$, which supports the findings of the previous sections that the effect of including or zeroing the corner modes washes out as redshift decreases; however, there are significant caveats.

One caveat is that for all of these profiles, it is important to note that the inner regions may not be sufficiently resolved to make strong comparisons between the simulations with and without corner modes. For one thing, at higher redshifts, where the differences in the power spectra are greater, there are no halos in our simulations with a sufficient number of particles to resolve the density profiles to the precision suggested by, e.g., \citet{Power2003}. Another caveat is that the small scales that exhibit differences due to the presence of corner modes (e.g. Figure~\ref{fig:pks}), probed by the inner regions of the halos, are smaller than the force softening of the simulations, $\epsilon = 0.04$\hmpc, below which the mass profile may not be converged~\citep{Power2003}. However, these convergence studies were performed for individual halos, not for the stacked halo profiles we show in the figures, which can perhaps be trusted more than individual halo profiles.

In any case, we find that for simulations at or below a resolution of $L/N=0.98$\hmpc, the effect of corner modes on halo profiles can be neglected.


\section{Conclusion}
\label{sec:disc}

A common practice in generating initial conditions of cosmological $N$-body simulations is to set the power spectrum to zero at $|\mathbf{k}|>$\kny, which can be non-zero off the axes of the simulation cube's window function (dubbed `corner modes'). We have run simulations with and without these corner modes to determine the effect of removing them on statistical measures of large-scale structure such as the power spectrum, density one-point PDF, halo mass function, and halo profiles.

We find that though the difference between the power spectra is largest at high redshift, even at $z=3$ the difference is at the level of 2\% above$\sim 80$\% of \kny\ and 0.5\% at $k=2$\ihmpc\ for a simulation with \kny$= 3.2$\ihmpc. These differences become smaller as $z \to 0$ because the corner modes are repopulated during the evolution of the simulation. Since comparing simulations with and without corner modes does not give us an idea of which is `right,' we also compare power spectra from simulations with the same particle resolution, but with a different resolution of initial modes, in order to determine whether including or removing these modes results in a better agreement between the low- and high-resolution simulations. To separate the effect of corner modes from that of particle resolution, we also compare power spectra of simulations with different particle resolution, but with the same initial modes. We find that including corner modes results in a better agreement between low- and high-resolution power spectra, but the effects of particle resolution are larger than the effects of the corner modes. 

The effect of removing corner modes is greatest at high redshift, high density, and small scales. Zeroing corner modes in initial conditions results in density distribution functions that are narrower with a suppressed high density tail. This difference at high densities is greatest at high redshift, reducing to $\sim 1\%$ at $z=6$. The variance of the 1\hmpc\ CIC overdensity PDF is suppressed by a factor of 2\% at $z=10$ when the corner modes are removed compared to when they are included, reducing to $< 1\%$ at $z=1$. On the other hand, there is no difference between the variance of 8\hmpc\ CIC overdensity PDFs. 

Though there are few halos at very high redshift in our simulations, removing corner modes in initial conditions does affect the halo mass function by 3\% at $z=6$ for the smallest halos, with fewer than 50 particles. We found no effect on the correspondence between mass functions in the low- and high-resolution simulations at $z<4$. Similarly, though removing corner modes does affect the inner regions of halo density profiles at high redshift, we did not find these differences to be significant.

Removing corner modes in the initial conditions results in a suppression of power below the Nyquist frequency and a reduced variance in the one-point density PDF until these corner modes are repopulated as the simulation evolves via nonlinear mode-mode coupling. The effect of corner modes is quite small, typically below the level of resolution and discreteness effects; it would be interesting to test the effect of removing corner modes using an initial conditions code that explicitly addresses discreteness effects~\citep[e.g.,][]{Garrison2016}. Subpercent effects will need to be considered for future surveys such as Euclid, DESI, WFIRST, and SKA.

\section*{Acknowledgements}

The authors are grateful to Adrian Jenkins, Stephane Colombi, and Donghui Jeong for useful discussions and comments on an earlier draft of this manuscript. 
BF acknowledges financial support from the Research Council of Norway (Programme for Space Research). 
NM acknowledges support from the European Research Council (DEGAS-259586) and the Science and Technology Facilities Council (ST/L00075X/1). 
MN was supported at IAP under the ILP LABEX (ANR-10-LABX-63) supported by French state funds managed by the ANR within the Investissements d'Avenir programme under reference ANR-11-IDEX-0004-02, and also by ERC Project No. 267117 (DARK) hosted by Universit Pierre et Marie Curie (UPMC) Paris 6, PI J. Silk. 
JW was supported by NSFC grants 11390372 and 11373029 and the Pilot-B project. 
MN and AS were supported at JHU by a grant in Data-Intensive Science from the Gordon and Betty Moore and Alfred P. Sloan Foundations.

\bibliographystyle{aasjournal}
\bibliography{cornermodes}

\end{document}